\documentclass[twocolumn]{aastex631}

\usepackage{graphicx}
\usepackage{amsmath,amssymb}
\usepackage{color}
\usepackage{units}
\usepackage{hyperref}
\usepackage{gensymb}
\usepackage{wrapfig,lipsum,booktabs}
\newcommand\um{$\,\rm{\mu}$m}
\newcommand\Lsun{L$_{\rm \odot}$}
\newcommand\Msun{M$_{\rm \odot}$}

\newcommand{\Hb}{H$\beta$}
\newcommand{\Pab}{Pa$\beta$}
\newcommand{\Pag}{Pa$\gamma$}
\newcommand{\Pad}{Pa$\delta$}
\newcommand{\Pae}{Pa$\epsilon$}
\newcommand{\Feii}{[Fe\textsc{ii}]$\lambda1.257$\um}
\newcommand{\Ha}{H$\alpha$}
\newcommand{\zs}{$z_{\rm spec}$}
\newcommand{\zp}{$z_{\rm phot}$}
\newcommand{\Hei}{He\textsc{i}}
\newcommand{\Oii}{[O\textsc{ii}]}

\newcommand{\Nii}{[N\textsc{ii}]}
\newcommand{\Siii}{[S\textsc{iii}]}
\newcommand{\Oiii}{[O\textsc{iii}]}

\newcommand\Msunyr{M$_{\rm \odot}$\,yr$^{-1}$}

\begin{document}

\title{RUBIES: JWST/NIRSpec resolves evolutionary phases of dusty star-forming galaxies at $z\sim2$}
\shorttitle{Dusty RUBIES}

\author[0000-0003-3881-1397]{Olivia R. Cooper}\altaffiliation{NSF Graduate Research Fellow}
\affiliation{Department of Astronomy, The University of Texas at Austin, Austin, TX, USA}
\affiliation{Cosmic Dawn Center (DAWN), Denmark}

\author[0000-0003-2680-005X]{Gabriel Brammer}
\affiliation{Cosmic Dawn Center (DAWN), Denmark}
\affiliation{Niels Bohr Institute, University of Copenhagen, Jagtvej 128, 2200 Copenhagen N, Denmark}

\author[0000-0002-9389-7413]{Kasper E. Heintz}
\affiliation{Cosmic Dawn Center (DAWN), Denmark}
\affiliation{Niels Bohr Institute, University of Copenhagen, Jagtvej 128, 2200 Copenhagen N, Denmark}
\affiliation{Department of Astronomy, University of Geneva, Chemin Pegasi 51, 1290 Versoix, Switzerland}

\author[0000-0003-3631-7176]{Sune Toft}
\affiliation{Cosmic Dawn Center (DAWN), Denmark}
\affiliation{Niels Bohr Institute, University of Copenhagen, Jagtvej 128, 2200 Copenhagen N, Denmark}

\author[0000-0002-0930-6466]{Caitlin M. Casey}
\affiliation{Department of Astronomy, The University of Texas at Austin, Austin, TX, USA}
\affiliation{Cosmic Dawn Center (DAWN), Denmark}

\author[0000-0003-4075-7393]{David J. Setton}\altaffiliation{Brinson Prize Fellow}
\affiliation{Department of Astrophysical Sciences, Princeton University, 4 Ivy Lane, Princeton, NJ 08544, USA}

\author[0000-0002-2380-9801]{Anna de Graaff}
\affiliation{Max-Planck-Institut f\"ur Astronomie, K\"onigstuhl 17, 69117 Heidelberg, Germany}

\author[0000-0002-3952-8588]{Leindert Boogaard}
\affiliation{Leiden Observatory, Leiden University, PO Box 9513, NL-2300 RA Leiden, The Netherlands}
\affiliation{Max-Planck-Institut f\"ur Astronomie, K\"onigstuhl 17, 69117 Heidelberg, Germany}

\author[0000-0001-7151-009X]{Nikko J. Cleri}
\affiliation{Department of Astronomy and Astrophysics, The Pennsylvania State University, University Park, PA 16802, USA}
\affiliation{Institute for Computational and Data Sciences, The Pennsylvania State University, University Park, PA 16802, USA}
\affiliation{Institute for Gravitation and the Cosmos, The Pennsylvania State University, University Park, PA 16802, USA}

\author[0000-0001-9885-4589]{Steven Gillman}
\affiliation{Cosmic Dawn Center (DAWN), Denmark}
\affiliation{DTU-Space, Technical University of Denmark, Elektrovej 327, DK-2800 Kgs. Lyngby, Denmark}

\author[0000-0003-0205-9826]{Rashmi Gottumukkala}
\affiliation{Cosmic Dawn Center (DAWN), Denmark}
\affiliation{Niels Bohr Institute, University of Copenhagen, Jagtvej 128, 2200 Copenhagen N, Denmark}

\author[0000-0002-5612-3427]{Jenny E. Greene}
\affiliation{Department of Astrophysical Sciences, Princeton University, 4 Ivy Lane, Princeton, NJ 08544, USA}

\author[0000-0002-4671-3036]{Bitten Gullberg}
\affiliation{Cosmic Dawn Center (DAWN), Denmark}
\affiliation{DTU-Space, Technical University of Denmark, Elektrovej 327, DK-2800 Kgs. Lyngby, Denmark}

\author[0000-0002-3301-3321]{Michaela Hirschmann}
\affiliation{Institute of Physics, Laboratory for Galaxy Evolution, Ecole Polytechnique Federale de Lausanne, Observatoire de Sauverny, Chemin Pegasi 51, 1290 Versoix, Switzerland}

\author[0000-0002-4684-9005]{Raphael E. Hviding}
\affiliation{Max-Planck-Institut f\"ur Astronomie, K\"onigstuhl 17, 69117 Heidelberg, Germany}

\author[0000-0003-3216-7190]{Erini Lambrides}\altaffiliation{NPP Fellow}
\affiliation{NASA-Goddard Space Flight Center, Code 662, Greenbelt, MD, 20771, USA}

\author[0000-0001-6755-1315]{Joel Leja}
\affiliation{Department of Astronomy and Astrophysics, The Pennsylvania State University, University Park, PA 16802, USA}
\affiliation{Institute for Computational and Data Sciences, The Pennsylvania State University, University Park, PA 16802, USA}
\affiliation{Institute for Gravitation and the Cosmos, The Pennsylvania State University, University Park, PA 16802, USA}

\author[0000-0002-7530-8857]{Arianna S. Long}
\affiliation{Department of Astronomy, The University of Washington, Seattle, WA USA}

\author[0000-0003-0415-0121]{Sinclaire M. Manning}
\altaffiliation{NASA Hubble Fellow}
\affiliation{Department of Astronomy, University of Massachusetts Amherst, 710 N Pleasant Street, Amherst, MA 01003, USA}

\author[0000-0003-0695-4414]{Michael V. Maseda}
\affiliation{Department of Astronomy, University of Wisconsin-Madison, Madison, WI 53706, USA}

\author[0000-0002-2446-8770]{Ian McConachie}
\affiliation{Department of Astronomy, University of Wisconsin-Madison, Madison, WI 53706, USA}

\author[0000-0002-6149-8178]{Jed McKinney}
\altaffiliation{NASA Hubble Fellow}
\affiliation{Department of Astronomy, The University of Texas at Austin, Austin, TX, USA}

\author[0000-0002-7064-4309]{Desika Narayanan}
\affiliation{Department of Astronomy, University of Florida, 211 Bryant Space Sciences Center, Gainesville, FL, USA}

\author[0000-0002-0108-4176]{Sedona H. Price}
\affiliation{Department of Physics and Astronomy and PITT PACC, University of Pittsburgh, Pittsburgh, PA 15260, USA}

\author[0000-0002-6338-7295]{Victoria Strait}
\affiliation{Cosmic Dawn Center (DAWN), Denmark}
\affiliation{Niels Bohr Institute, University of Copenhagen, Jagtvej 128, 2200 Copenhagen N, Denmark}

\author[0000-0001-8928-4465]{Andrea Weibel}
\affiliation{Department of Astronomy, University of Geneva, Chemin Pegasi 51, 1290 Versoix, Switzerland}

\author[0000-0003-2919-7495]{Christina C.\ Williams}
\affiliation{NSF's National Optical-Infrared Astronomy Research Laboratory, 950 North Cherry Avenue, Tucson, AZ 85719, USA}
\affiliation{Steward Observatory, University of Arizona, 933 North Cherry Avenue, Tucson, AZ 85721, USA}

\date{\today}

\begin{abstract}

The dearth of high quality spectroscopy of dusty star-forming galaxies (DSFGs) --- the main drivers of the assembly of dust and stellar mass at the peak of activity in the Universe --- greatly hinders our ability to interpret their physical processes and evolutionary pathways. We present JWST/NIRSpec observations from RUBIES of four submillimeter-selected, ALMA-detected DSFGs at cosmic noon, $z\sim2.3-2.7$. While photometry uniformly suggests vigorous ongoing star formation for the entire sample in line with canonical DSFGs, the spectra differ: one source has spectroscopic evidence of an evolved stellar population, indicating a recent transition to a post-starburst phase, while the remainder show strong spectroscopic signatures of ongoing starbursts. All four galaxies are infrared-luminous (log$_{10}$\,$L_{\rm{IR}}$/\Lsun\ $>12.4$), massive (log$_{10}\,M_\star$/\Msun\ $>11$), and very dust-obscured ($A_V\sim3-4$ ABmag). Leveraging detections of multiple Balmer and Paschen lines, we derive an optical attenuation curve consistent with Calzetti overall, yet an optical extinction ratio $R_V\sim2.5$, potentially indicating smaller dust grains or differences in star-dust geometry. This case study provides some of the first detailed spectroscopic evidence that the DSFGs encompass a heterogeneous sample spanning a range of star formation properties and evolutionary stages, and illustrates the advantages of synergistic JWST and ALMA analysis of DSFGs.

\end{abstract}

\keywords{Submillimeter astronomy (1735) --- Starburst galaxies (1570) --- Galaxy spectroscopy (2171) --- Interstellar dust (836)}

\section{Introduction}
\label{sec:intro}

Despite the accumulating wealth of multi-wavelength data over wide-area extragalactic fields, our understanding of the dust-obscured Universe remains largely limited to relatively few photometric data points, or for the fortuitous subset of dusty galaxies, low significance spectra. The bulk of star formation in the Universe is generated in dusty star-forming galaxies \citep[DSFGs;][]{2014madaudickinson}, which are characterized by high star formation rates ($\gtrsim100\,$\Msunyr), large stellar masses ($\gtrsim10^{10}\,$\Msun), and massive dust reservoirs ($\gtrsim10^{8}\,$\Msun) \citep*[see][for a review]{2014casey,2020hodge}. The vast majority of DSFGs --- given their low on-sky source density of $\lesssim10^{4}$\,deg$^{-2}$ \citep{2016oteo,2020bethermin,2020lagos} --- are discovered within wide-area, single-dish submillimeter surveys, which have identified thousands of DSFGs via their strong dust emission in the submillimeter \citep[e.g.][]{1997smail,1998hughes,2005chapman,2008scott,2010pilbratt,2010eales,2011aretxaga,2012oliver,2017koprowski,2019simpson}. A subset of these single-dish sources have been followed up interferometrically, which provides accurate pinpointing of the often blended objects \citep[e.g.][]{2004chapman,2005chapman,2013karim,2013hodge,2015dacunha,2019stach}. Multi-wavelength follow-up introducing additional photometric constraints in the optical/near-infrared (OIR) has suggested that the majority of DSFGs sit at $1<z<3$ \citep[][]{2005lefloch,2011magnelli,2012casey_zsurvey,2013magnelli}, bracketing the peak of cosmic star formation, and indeed dominating cosmic star formation overall \citep{2013gruppioni,2014casey}. However, even when these highly attenuated, submillimeter-selected galaxies are detected photometrically in the OIR, due to their dust obscuration, we often lack necessary information to even break degeneracies with redshift, let alone gain any detailed understanding of the physics regulating these extreme sites of star formation. In other words, the dearth of spectroscopy for DSFGs greatly hinders our ability to interpret their star formation properties in detail. Consequently, our understanding of the evolutionary pathways of the primary players in the buildup of stellar mass at the peak of activity in the Universe is limited.

A handful of studies in the past decade have worked towards spectroscopic confirmation of DSFGs, but given historically imprecise or inaccurate photometric redshifts for DSFGs, in addition to often uncertain OIR counterpart identification, the observations returned fairly low spectroscopic yields. General-purpose OIR spectroscopic campaigns over $\gtrsim0.5\,$deg$^2$ within legacy fields \citep[e.g. UDSz;][]{2016maltby}, which are generally efficient for typical star-forming galaxies up to cosmic noon, are mostly unsuccessful at measuring spectroscopic redshifts for such dust-attenuated and rare objects as DSFGs, much less obtaining multi-line information. Dedicated spectroscopic follow-up of DSFGs targeting rest-frame ultraviolet/optical emission lines have managed to measure spectroscopic redshifts for some DSFGs, with some bias towards DSFGs that are optically bright, and potentially towards those containing active galactic nuclei \citep[AGN; e.g.][]{2003chapman,2003smail,2005chapman}. Deep spectroscopic surveys of DSFGs in the rest-frame optical have also faced challenges given inaccurate or non-existent photometric redshifts for DSFGs, and as a whole have returned relatively low spectroscopic yields \citep[e.g.][]{2017casey,2017danielson}, with many sources only detected via single emission lines. In the cases where rest-frame near-infrared spectra were obtained for DSFGs, in the majority of cases these data were ground-based and heavily skyline dominated. Faced with technical limitations observationally, and given their rarity, OIR counterpart uncertainty and faintness, and characteristically large photometric redshift uncertainty, obtaining high quality and detailed OIR spectra of DSFGs was not possible pre-JWST.

By pushing deeper and to longer near-infrared wavelengths, JWST has proven its efficacy at detecting obscured, high redshift objects like DSFGs \citep[e.g.][]{2022chenb,2022cheng,2023gillman,2023mckinney,2023rujopakarn,2023wu,2023zavala,2023xiao,2023herard-demanche,2024gillman}. Observations with NIRSpec/PRISM afford the wavelength coverage and resolution to capture multiple rest-frame OIR emission lines for DSFGs. For example, \citet{2023barrufet} target a sample of red galaxies at $z>3$ and detect multiple spectroscopic features of the galaxies, many of which are likely DSFGs or their slightly less extreme cousins that are missed by shallow submillimeter surveys due to their relative submillimeter-faintness. The NIRSpec/MSA survey, Red Unknowns: Bright Infrared Extragalactic Survey \citep[RUBIES, GO\#4233, PIs A. de Graaff and G. Brammer;][]{2024degraaff}, is designed to achieve spectroscopic completeness for bright, red objects at cosmic noon and earlier. Target selection and mask design is optimized to return high completeness in particular for the most extreme sources in this color space, which includes DSFGs. At the redshift regime where most DSFGs sit ($1<z<3$), RUBIES covers multiple Paschen and Balmer lines, in addition to other species emerging from nebular regions. Joined with far-infrared constraints from ALMA, these data enable robust measurements of nebular, stellar, and dust properties, as well as direct constraints on ongoing star formation activity and history.

In this paper, we present JWST NIRSpec/PRISM observations of four submillimeter-selected, ALMA-continuum-confirmed sources \citep[from AS2UDS;][]{2019stach} at $z\sim2.5$ targeted with RUBIES. These observations represent some of the first high quality rest-frame OIR spectra of DSFGs, and for the first time provide spectroscopic insight into the heterogeneous nature of the DSFG population, direct evidence of evolutionary phases of DSFGs, and detailed constraints of their stellar and dust content. We describe the sample and observations in \textsection 2, and in \textsection 3 we present analysis of physical characteristics based on multi-wavelength spectroscopic and photometric data. \textsection 4 discusses the implications of our measurements, and \textsection 5 summarizes. We assume a Chabrier IMF \citep{2003chabrier} and \textit{Planck} cosmology throughout this paper, adopting $H_0 = 67.7 \rm{\,km\,s^{-1}\,Mpc^{-1}}$ and $\Omega_{\lambda} = 0.6911$ \citep{planck}.

\section{Sample \& Observations}

\subsection{Sample selection and spectroscopic data}

JWST spectroscopic data for our sample were obtained by RUBIES \citep[GO\#4233, PI: A. de Graaff and G. Brammer,][]{2024degraaff}, a JWST Cycle 2 program using the NIRSpec microshutter array (MSA) targeting galaxies in the Cosmic Assembly Near-infrared Deep Extragalactic Legacy Survey \citep[CANDELS;][]{candels} regions of the Extended Groth Strip (EGS) and Ultra Deep Survey (UDS) fields. The primary science targets for RUBIES are galaxies detected in F444W from JWST/NIRCam imaging in the Cosmic Evolution Early Release Science Survey \citep[CEERS, ERS\#1345, PI: S. Finkelstein,][]{2022finkelstein,ceers}\footnote{Available for download at \href{ceers.github.io/releases.html}{ceers.github.io/releases.html} and on MAST as High Level Science Products via \href{doi:10.17909/z7p0-8481}{doi:10.17909/z7p0-8481} \citep{ceers}}, and the Public Release IMaging for Extragalactic Research Cycle 1 program \citep[PRIMER, GO\#1837, PI: J. Dunlop,][]{primer}. RUBIES is optimized to reach high spectroscopic completeness for bright, red sources at $z > 3$. We selected targets from the DAWN JWST archive (DJA) \texttt{grizli} catalogs, using imaging reduction version v7.2 \citep{2023grizli}, and adopted spectra from the DJA \citep{2024heintz}. Specifically, priority 1 (and 2) targets have F150W$-$F444W $>3(2)$ and F444W $< 27$. Full description of the RUBIES survey is presented in \citet{2024degraaff}.

To select our sample, we search the initial three RUBIES pointings in UDS (i.e. all RUBIES data in ALMA-accessible fields as of our DJA archival search) for crossmatches to archival ALMA sources (within $1''$), to identify known DSFGs observed by RUBIES. The sample includes all known crossmatches as of 2024 September 13. From this search, we identify five ALMA-bright sources, four at cosmic noon ($z\sim2.5$), and one at higher redshift ($z\sim4.5$, which will be presented in a later paper; O. Cooper et al. in prep.). All DSFGs in our sample were ALMA-confirmed via the ALMA SCUBA-2 UDS survey \citep[AS2UDS; ][]{2015simpsonb,2017simpson,2019stach}, and were initially discovered as submillimeter sources by the SCUBA-2 Cosmology Legacy Survey \citep[S2CLS; ][]{s2cls}; the submillimeter data are described further in \textsection2.3. Within the three RUBIES masks in UDS, there were only two known DSFGs that fell into the NIRSpec field of view but were not targeted; in other words, approximately $70\%$ of the submillimeter-bright galaxies in the field of view were observed. The two non-targeted DSFGs (AS2UDS.0141.0 \& AS2UDS.0152.0) are bright and red in NIRCam, but are slightly bluer in F150W$-$F444W than the targeted sample overall, and have low photometric redshifts of $z<2$ (with uncertainty $\Delta z \sim0.2$), which further decreased their priority weighting for target selection. For this work, we hone in on the subsample of four NIRSpec-observed DSFGs at cosmic noon, which are all classified photometrically as typical DSFGs by \citet{2020dudz}, each with $S_{870\mu\rm m}>1$\,mJy.

The four galaxies presented here were observed on 2024 January 16, 18, \& 19 over three masks in the UDS field. The MSA pointings were observed for 48 minutes each in the PRISM/CLEAR and the G395M/F290LP spectroscopic modes, with a standard 3-shutter slitlet and 3-point nodding pattern. For the targets observed on multiple masks, we adopt the observation with a higher S/N for our analysis.

The NIRSpec spectra were reduced and extracted with version 3 of \texttt{msaexp} \citep{2023brammerb}. Uncalibrated NIRSpec exposures are processed through the Stage 1 steps of the JWST calibration pipeline, inserting an improved mask for ``snowball'' artifacts using \texttt{snowblind} \citep{2024davies} and applying a $1/f$ correction. Then, individual slits are identified and the 2D unrectified data are flat fielded. Background is subtracted by taking image differences of the 2D flux-calibrated spectra cut out from the exposures taken at three offset positions within the standard nod sequence. The 1D spectra are optimally extracted \citep{1986horne} from the rectified 2D spectra \citep[see][for further details]{2024heintz}. Lastly, we apply wavelength dependent slit loss corrections (resulting in a flux calibration factor of $\sim1-2\times$) anchored to the NIRCam photometry to derive flux-calibrated 1D spectra.

\subsection{Optical to near-infrared photometric data}

Our targets were selected for RUBIES by public JWST/NIRCam imaging from the PRIMER survey, which obtained NIRCam imaging in 8 bands (F090W, F115W, F150W, F200W, F277W, F356W, F444W and F410M) and MIRI imaging in the F770W and F1800W bands. We use latest version (v7.2) of the publicly available JWST image mosaics from the DJA, which were reduced using \texttt{grizli} \citep{2023grizli}; see also \citet{2023valentino} for a description. We also include archival Hubble Space Telescope (HST) imaging from the CANDELS survey \citep[][]{2011grogin,2011koekemoer}, which obtained imaging in F435W, F606W, F814W, F105W, F125W, F140W, and F160W.

The flux measurement routine is described in detail in \citet{2024weibel}. Briefly, to measure the NIRCam photometry we use \texttt{SourceExtractor} \citep{1996bertin} in dual image mode, with an unconvolved, inverse-variance weighted image stack in F277W, F356W and F444W taken as the detection image. Then, we construct a mosaic for each NIRCam band, point-spread-function (PSF) matched to the F444W image. From these mosaics, fluxes are measured in circular apertures of radius ${0}\farcs{16}$. These aperture fluxes are then scaled to the flux measured through a Kron aperture on a PSF-matched version of the detection image. We scale to the total flux by dividing by the encircled energy of the Kron aperture on the F444W PSF. MIRI fluxes were then measured within $0\farcs5$ apertures, and matched to F444W resolution by multiplying the measured MIRI flux by the ratio of the encircled energies at ${0}\farcs{5}$ of the F444W and the MIRI PSF, respectively.

\subsection{Far-infrared data}

Submillimeter data for our sample were obtained by the ALMA SCUBA-2 UDS survey \citep[AS2UDS; ][]{2015simpsonb,2017simpson,2019stach}. AS2UDS is a complete ALMA follow-up of all bright single-dish sources (detected at $>4\sigma$, or equivalently, $S_{850\,\mu\rm m} \geq 3.6$\,mJy) selected from the SCUBA-2 Cosmology Legacy Survey \citep[S2CLS; ][]{s2cls} of the UKIRT Infrared Deep Sky Survey (UKIDSS) UDS field. The SCUBA-2 survey covers $0.96$\,deg$^2$ of the UDS field, with a median depth of $\sigma_{850} = 0.88$\,mJy\,beam$^{-1}$, and varying depth across the field overall below 1.3 mJy \citep{2017geach}. AS2UDS observations were obtained in Cycles 1, 3, 4, and 5 in Band 7 (870\,\um{}) across the S2CLS-UDS field, with a median survey depth of $0.3^{+0.1}_{-0.2}$\,mJy\,beam$^{-1}$. Each of our sources is detected at $>4\sigma$ with ALMA Band 7; we adopt flux densities presented in \citet{2020dudz}, which compiles multiwavelength photometry spanning ultraviolet to radio for the 707 AS2UDS sources.

To constrain the dust peak of the far-infrared spectral energy distribution (FIR SED), we include Herschel/Spectral and Photometric Imaging Receiver (SPIRE) data at 250\um{}, 350\um{}, and  500\um{} from the Herschel Multi-tiered Extragalactic Survey \citep[HerMES; ][]{2012oliver}. Note that all of our sources are non-detected in Herschel/Photoconductor Array Camera and Spectrometer (PACS) 100\um{} and 160\um{} imaging and in Spitzer/Multiband Imaging Photometer (MIPS) 24\um{} imaging. We adopt SPIRE flux densities from the multiwavelength catalog for AS2UDS presented in \citet{2020dudz}. These measurements are deblended following the same procedure as \citet{2014swinbank}, wherein the ALMA sources are used as positional priors in the deblending of the SPIRE maps. However, as the data are confusion limited, we adopt the confusion error from \citet{spire} for any sources with derived errors less than these confusion error limits. This is a conservative estimate of the noise given the uncertainty in positional priors of deblended catalogs.

\subsection{X-ray \& Radio data}

We check for X-ray detections for our sample within the deep \textit{Chandra} map of the UDS field obtained by the X-UDS survey, with a 1$''$ tolerance \citep{2018kocevski}. The X-UDS survey covers $0.33$\,deg$^2$ of the UDS field, with a deeper central component; the average exposure time per pixel is 600\,ks in the deep central region and 200\,ks in the outer region. The catalog of X-ray point sources has a detection threshold of $4.4\times10^{-16}$\,erg\,s$^{-1}$\,cm$^{-2}$ in the full band ($0.5-10$\,keV). All of our sources fall into the \textit{Chandra} footprint, and none are detected \citep[see ][]{2019stach}.

We adopt radio flux densities from AS2UDS \citep[][]{2020dudz}, which are derived from 1.4\,GHz imaging taken with the Very Large Array (VLA), and have an average depth of $10$\,$\mu$Jy beam$^{-1}$. Half of our sample is detected at 1.4\,GHz as sub-mJy sources, and the other two sources are not detected above $4\sigma$.

\section{Analysis \& Results}

In this work, we set out to measure and compare star formation rate indicators across energy regimes and timescales, as well as any other accessible tracers of stellar and dust content for our sample of DSFGs. 

\subsection{Spectroscopic Measurements}

\begin{figure*}[]
    \centering
    \includegraphics[angle=0,trim=0in 0in 0in 0in, clip, width=1.\textwidth]{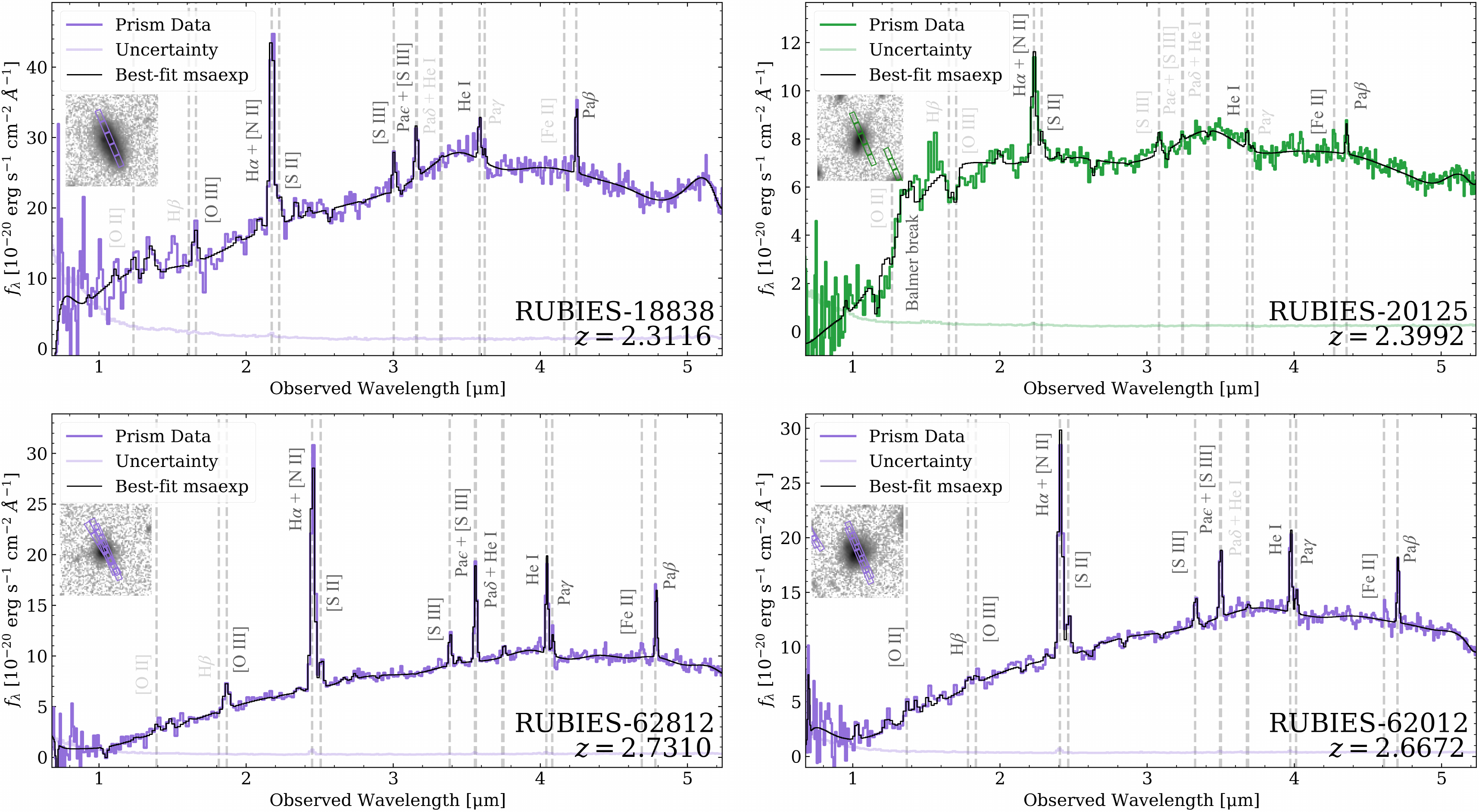}
    \caption{JWST/NIRSpec 1D PRISM spectra and uncertainty for our sources (green/purple and light green/purple, respectively) and best-fit model from \texttt{msaexp} (black). We present the spectra in two different colors to emphasize the distinction in their star-formation properties, using green for the more evolved, post-starburst galaxy, and purple for the strongly starbursting galaxies. Measured spectroscopic redshifts and object names are listed for each source. Wavelengths of redshifted spectroscopic features are shown as dashed vertical lines, with detected lines or features (S/N$>3$) labeled in dark gray and non-detected lines labeled in light gray for reference. The JWST/NIRCam F444W cutout for each source is shown as an inset ($2''$ wide), with the position of the NIRSpec slit overplotted on each image.}
    \label{fig:spec}
\end{figure*}

The PRISM spectra of the sources generally demonstrate bright, red continuua, and strong detections of multiple rest-frame OIR emission lines. Two sources (RUBIES-62812 and RUBIES-62012) featured 10+ emission line detections, including significant Balmer and Paschen emission. One source (RUBIES-18838) has a complex, multi-component spectrum, with significant \Ha\ emission but generally fewer strong lines. The final source (RUBIES-20125) shows a strong Balmer break, a flatter continuum slope, and fewer emission lines. The source properties, including the spectroscopic features, are described in more detail in \textsection3.5.

Spectroscopic redshifts were measured with \texttt{msaexp} based on detected spectral features, either via the G395M medium resolution spectra when strong lines were detected within the wavelength coverage, or from the PRISM spectra, which affords a broader wavelength coverage but lower precision. Redshifts for RUBIES-62812 and RUBIES-62012 were derived from multiple lines within the G395M data (from detections of \Siii, \Pae, \Pad, \Hei, \Pag, \Feii, and \Pab). No strong lines were detectable in G395M for RUBIES-20125 and RUBIES-18838 given their redshifts and generally weaker emission lines at observed-frame $3-4$\,\um{}; redshifts for these sources were derived via multiple lines detected in the PRISM data. A reliable redshift was accessible in every case due to at least two high signal-to-noise line detections (S/N$>3$). This G395M or PRISM derived spectroscopic redshift solution was then fixed and input into \texttt{msaexp} to fit the deep PRISM spectra, which utilizes a library of spectral lines and cubic splines to find a flexible line plus continuum model. Emission line fits are generated pixel-integrated Gaussians, with widths anchored to the PRISM wavelength-dependent spectral resolution curve; note that PRISM does not spectrally resolve emission lines, except in cases where emission lines are very broad due to e.g. broad-line AGN. Note that \Feii\ is not included in the \texttt{msaexp} templates, so the feature is not fit in the model, though it is detected for three out of four targets. Additionally, the contribution from stellar absorption is not accounted for in the emission line fitting, therefore the inferred emission may be underestimated; in particular, this may impact the measurement of \Hb, which is marginally detected in only one galaxy (RUBIES-62012). As we discuss in \textsection3.2, we anchor our nebular attenuation measurements primarily to the strongly detected lines, namely \Pab\ and \Ha, therefore we expect minimal impact on our results. The PRISM spectra and best-fit \texttt{msaexp} models are shown in Figure \ref{fig:spec}, and line fluxes and associated errors taken from the model fit to the spectrum are presented in Table \ref{tab:spec}.

\begin{deluxetable*}{lcccccccccc}[t]
\setlength{\tabcolsep}{0.02in}
\tablewidth{1.\textwidth}
\tabletypesize{\footnotesize}
\tablecaption{Spectroscopic Properties\label{tab:spec}} 
\tablehead{
\colhead{Name} & {R.A.} & {Dec.} & {\zs}  & {$F_{\rm H\alpha+[NII]}$} & {EW$_{\rm H\alpha+[NII]}$} & {$F_{\rm H\beta}$} & {$F_{\rm Pa\beta}$} & {$A_V$} & {SFR$_{\rm H\alpha}$} & {SFR$_{\rm Pa\beta}$}\\
	\colhead{}  & \colhead{J2000} & \colhead{J2000}  & \colhead{} & \colhead{$10^{-20}\rm{\,erg\,s^{-1}\,cm^{-2}}$} & \colhead{$\rm\AA$} & \colhead{$10^{-20}\rm{\,erg\,s^{-1}\,cm^{-2}}$} & \colhead{$10^{-20}\rm{\,erg\,s^{-1}\,cm^{-2}}$} & \colhead{ABmag} & \colhead{\Msunyr} & \colhead{\Msunyr} \\
	\colhead{(1)}  & \colhead{(2)} & \colhead{(3)} & \colhead{(4)} & \colhead{(5)} & \colhead{(6)} & \colhead{(7)} & \colhead{(8)} & \colhead{(9)} & \colhead{(10)} & \colhead{(11)}} 
\startdata
RUBIES-18838 & 02:16:57.61 & --05:16:56.06 & 2.3116 & $11128\pm480$ & 209 & $<320$ & $1927\pm191$ & $3.8\pm2.1$ & $220\pm130$ & $220\pm130$ \\ 
RUBIES-20125  & 02:16:58.02 & --05:16:46.83 & 2.3992 & $1179\pm117$ & 63 & $-553\pm140$ & $202\pm48$ & $4.0\pm2.2$ & $50\pm30$ & $50\pm30$ \\ 
RUBIES-62812  & 02:17:02.93 & --05:12:12.50 & 2.7310 & $7865\pm185$ & 321 & $<120$ & $1294\pm70$ & $3.1\pm2.1$ & $310\pm180$ & $320\pm180$  \\
RUBIES-62012 & 02:17:03.02 & --05:12:18.34 & 2.6672 & $6834\pm198$ & 204 & $629\pm156$ & $1140\pm74$ & $3.6\pm2.1$ & $280\pm160$ & $290\pm160$ \\
\enddata
\tablecomments{Columns: (1) Object ID, (2) Right ascension, (3) Declination, (4) spectroscopic redshift derived from multiple emission lines identified in NIRSpec/G395M, (5) observed H$\alpha+$\Nii\ emission line flux and uncertainty, (6) rest-frame equivalent width (EW) of H$\alpha+$\Nii, (7) observed H$\beta$ emission line flux and uncertainty or measured upper limit, (8) observed Pa$\beta$ emission line flux and uncertainty, (9) attenuation $A_V$ measured from the slit-loss corrected Paschen-Balmer decrement, (10) dust-corrected star formation rate measured from the slit-loss corrected H$\alpha+$\Nii\ emission line luminosity, and (11) dust-corrected star formation rate measured from the slit-loss corrected Pa$\beta$ emission line luminosity.}
\end{deluxetable*}

\subsection{Attenuation Curve and Dust-corrected Spectroscopic Properties}

A nebular attenuation curve can be derived directly from a spectrum using detections of multiple Balmer and/or Paschen lines. To best inform our spectroscopic measurements of star formation and dust attenuation properties, we derive an attenuation curve from all detected (S/N $>3$) Paschen and Balmer emission line pairs. Given the small sample and in particular, the relative dearth of detected emission lines from RUBIES-18838 \& RUBIES-20125, construction of the attenuation curve is predominantly driven by the galaxy with the largest number of detected emission lines, RUBIES-62012. We implement the Balmer-line-ratio-based method described in \citet{2020reddy}, but anchor our derivation to our reddest detected line, \Pab\ (12818\,\rm\AA), whereas they use Balmer-series lines only and anchor to \Ha, the reddest line in their data. This method has also been implemented in other recent works \citep[e.g.][]{2022prescott,2022cleri,2023reddy,2024sanders}. Using the slit-loss corrected spectra, we calculate the observed versus intrinsic ratio of the \Pab\ line flux to all other detected Hydrogen emission lines (\Pag, \Pad, \Ha, and \Hb) to construct the attenuation curve.

Given the low resolution of the PRISM spectra ($R\sim100$), \Ha\ and \Nii\ are blended, so the contribution of \Nii\ to our measured \Ha+\Nii\ line flux is unconstrained. Note that the medium resolution G395M wavelength coverage is too red to capture the \Ha\ line for our sources. To account for some \Nii\ emission, we allow the \Nii/\Ha\ fraction to vary. While two out of four sources show evidence of AGN in the radio (see subsections on individual sources under \textsection3.5), we do not detect broad line signatures in the rest-frame optical spectra. Nevertheless, we cannot constrain the contribution from AGN versus star formation to the \Ha+\Nii\ line flux, or the relative contribution of \Nii\ and \Ha\ in the blended line. Therefore, we assume an uncertain \Nii/\Ha\ ratio term of $0.5\pm0.5$, to allow for uncertainty in the contribution from \Nii, informed by the typical fraction for high mass galaxies at cosmic noon \citep{2015shapley}. Despite the unconstrained contribution of \Nii, we note that adjusting the \Nii\ fraction has minimal impact on the attenuation curve fit given constraints from the non-\Ha\ Paschen-Balmer line ratios, and has greater implications for the Paschen-Balmer-decrement-based $A_V$ measurements for individual sources: a higher contribution of \Nii\ increases the measured $A_V$ for a given source. This uncertainty is included in the reported errors for our spectroscopic-based $A_V$ and dust-corrected \Ha\ and \Pab\-based SFRs, listed in Table \ref{tab:spec}.

Based on the slit-loss corrected emission line fluxes, we begin constructing the attenuation curve by finding $A^\prime(\lambda)$, defined here as the attenuation in magnitudes relative to \Pab, which can be expressed as:

\begin{equation}
\begin{split}
    A^\prime(\lambda) = 2.5 \left[\log_{10}\left(\frac{F_{\rm obs}(\rm Pa \beta) / F_{\rm obs}(\lambda)}{F_0(\rm Pa \beta) / F_0(\lambda)} \right)\right] + 1,
    \label{eq:ap}
\end{split}
\end{equation}

where $F_{\rm obs}(\rm Pa \beta)/F_{\rm obs}(\lambda)$ is the observed ratio of the \Pab\ line flux to the observed line flux of each Balmer/Paschen line (\Pag, \Pad, \Ha, and \Hb), and $F_0(\rm Pa \beta)/F_0(\lambda)$ is the intrinsic ratio of line fluxes for each emission line pair. For these intrinsic ratios, we assume Case B recombination, $T=10,000\,$K, and $n_e = 100\,\rm cm^{-3}$ \citep{2006osterbrock}. $A^\prime(\lambda)$ is offset from $A(\lambda)$ by the wavelength-independent normalization constant $[A(\rm Pa\beta)-1]$. These $A^\prime(\lambda)$ points are then fit to a function linear in $1/\lambda$, weighted by the line flux errors. 

From the best-fit line to $A^\prime(\lambda)$, the attenuation curve $k^\prime(\lambda)$ can be calculated based on the definition of $k(\lambda) = A(\lambda)/E(B-V)$, where

\begin{equation}
    k^\prime(\lambda) = \frac{A^\prime(\lambda)}{A^\prime(4400\rm\AA) - A^\prime(5500\rm\AA)}
    \label{eq:kp}
\end{equation}

for the color excess $E(B-V)$ defined at $\lambda_B = 4400\,\rm\AA$ and $\lambda_V = 5500\,\rm\AA$. The errors on line fluxes are propagated throughout, and we normalize the functional form of the attenuation curve such that $k^\prime(\rm Pa \beta) = 1$ for comparison to other attenuation curves in the literature (see \textsection 4.1).

The final functional form of the derived nebular dust attenuation curve is normalized such that at sufficiently long wavelengths, $k(\lambda)=0$. Here, we set the curve to zero at $\lambda=28000\,\rm\AA$, consistent with \citet{2015reddy} and for attenuation curves derived for the Milky Way, LMC, SMC, and local star-forming galaxies \citep[][respectively]{1989cardelli,2003gordon,2000calzetti}. 

While we assume this normalized attenuation curve $k(\lambda)$ throughout, we calculate  $A_V(\lambda=5500\,\rm\AA)$ for each galaxy in our sample by calculating $E(B-V)$ based on each galaxy's Paschen-Balmer decrement. In other words, $k(\lambda)$ is multiplied by $E(B-V)$ to find $A_V(\lambda=5500\,\rm\AA)$. Explicitly, this color excess is found by the following: 

\begin{equation}
\begin{split}
    E(B-V) &= \frac{2.5}{k(\rm Pa\beta) - k(\rm H\alpha)} \\
    &\times \log_{10} \left[\frac{F_{\rm obs}(\rm H \alpha)/F_{\rm obs}(\rm Pa \beta)}{F_0(\rm H \alpha)/F_0(\rm Pa \beta)}\right],
    \label{eq:ebv}
    \end{split}
\end{equation}

where $F_0(\rm H \alpha)/F_0(\rm Pa \beta) = 17.39$ \citep{2006osterbrock}. The spectroscopically-derived $A_V$ for each galaxy is reported in Table \ref{tab:spec}, with errors propagated from both measured line flux uncertainties and from our assumed \Nii\ fraction of $0.5\pm0.5$.

The normalized functional form of $A(\lambda)$ also enables dust-corrected star formation rate (SFR) measurements based on slit-loss corrected \Ha\ and \Pab\ line luminosities. For each galaxy, we measure SFR$_{\rm H\alpha}$ and SFR$_{\rm Pa\beta}$ using the empirical relations in \citet{2012kennicuttevans}, using calibrators from \citet{2011hao,2011murphy}. We report the estimated these dust-corrected SFRs with errors propagated from measured line flux uncertainties in Table \ref{tab:spec}.

\subsection{Far-infrared Properties}

As part of the AS2UDS program, \citet{2020dudz} fit the full ultraviolet-to-radio SED (with pre-JWST photometry) using the energy balance procedure \texttt{MAGPHYS}; here we fit the FIR SED to verify the derived FIR properties independently of the OIR photometry. Each galaxy's FIR SED is fit to a modified blackbody added piecewise with a mid-infrared power law fit using Bayesian analysis tool \texttt{MCIRSED} \citep{2022drew}; best-fit SEDs are derived based on a Markov chain Monte Carlo (MCMC)
convergence. The mid-infrared power law is joined to the modified blackbody at the point where the blackbody slope is equal to the power law index $\alpha_{\rm MIR} = 2$ \citep[consistent with other works, e.g.][]{2010kovacs,2012casey,2012u}. The general opacity model is assumed, where the optical depth ($\tau$) equals unity at $\lambda_{\rm rest} = 200$\um{} \citep[e.g.][]{2011conley,2012greve}, though the exact adopted value of the opacity transition has little impact on the modeled SED.

Input into the FIR SED fitting routine is the spectroscopic redshift and the FIR/mm photometry (at 250\um{}, 350\um{}, 500\um{}, and 870\um{}) with associated uncertainties for the flux density measurements. Our fixed parameters are $\alpha_{\rm MIR} = 2$, $\beta=2$, $\lambda_0 = 200$\um{} (the wavelength where optical depth equals unity, near the intrinsic peak of the dust SED). Note that we fix the dust emissivity spectral index ($\beta$); while this index has been suggested to vary for submillimeter sources \citep[e.g. ][]{2021dacunha,2022cooper}, we opt to fix to $\beta=2$ for our fits given our lack of constraint on the Rayleigh-Jeans tail. As none of these parameters can be directly constrained in this dataset, these broad population-averaged values are assumed. 

Using \texttt{MCIRSED}, we find the best-fit dust SED with measurements for each of the following free parameters: total IR luminosity ($L_{\rm IR}$, taken from 8 to 1000 \um), dust temperature ($T_{\rm dust}$), and rest-frame peak wavelength ($\lambda_{\rm peak}$). The last two variables have a fixed relationship given our assumed opacity model with $\lambda_0 = 200$\,\um{} (see Figure 20 of \citealt*[][]{dsfgcaseyreview}). We then convert our $L_{\rm IR}$ measurements to a SFR$_{\rm IR}$ using the TIR calibrator from \citet{2011hao} and \citet{2011murphy}, as summarized in the \citet{2012kennicuttevans} review. 

To derive a dust mass for each galaxy, we assume the optically thin, isotropic case wherein dust emission emerges as a simple, single-temperature modified blackbody ($B_\nu$). We adopt a single mass-weighted dust temperature of 25\,K \citep[consistent with][]{2016scoville,2019casey}, given that the vast majority of a galaxy's dust mass radiates at cold temperatures, though hot dust can contribute substantially to a galaxy’s IR luminosity (reflected in measurements of the dust temperature from SED fitting, which are instead luminosity-weighted). Converting the observed flux density $S_{870\mu m}$ into luminosity ($L_{870\mu m}$), and adopting a mass absorption coefficient of dust from \citet{2001li} at the rest-frame frequency ($\kappa_\nu$), we can estimate a dust mass via the following equation:

\begin{equation}
    M_{\rm dust} = \frac{L_{870\mu m}}{4 \pi \kappa_\nu B_\nu(\rm{T}=25\,\rm{K})}
\end{equation}

This correlation with $S_{870\mu m}$ and $M_{\rm dust}$ has been well established, wherein the 870\um\ flux density traces cold dust mass \citep[e.g.][]{2014scoville,2018liang,2020dudz}. Indeed, with this method, we find consistent dust masses within uncertainties as those reported in \citet{2020dudz} for our sample. The derived FIR properties are listed in Table \ref{tab:fir}.

\begin{deluxetable}{lccccc}
\setlength{\tabcolsep}{0.01in}
\tablewidth{0.8\textwidth}
\tabletypesize{\footnotesize}
\tablecaption{Far-infrared Properties\label{tab:fir}} 
\tablehead{
\colhead{Name} & {AS2UDS ID} & {$S_{870\mu\rm m}$}  & {log$_{10}$$(L_{\rm{IR}})$} & {SFR$_{\rm IR}$} & {log$_{10}$($M_{\rm dust}$)}\\
	\colhead{}  & \colhead{} & \colhead{mJy}  & \colhead{\Lsun} & \colhead{\Msunyr} & \colhead{\Msun}\\
	\colhead{(1)}  & \colhead{(2)} & \colhead{(3)} & \colhead{(4)} & \colhead{(5)} & \colhead{(6)}} 
\startdata
RUBIES-18838 & 258.1 & $1.12\pm0.18$ & $12.4_{-0.4}^{+0.2}$ & $350^{+240}_{-220}$ & $8.56\pm0.07$\\ 
RUBIES-20125  & 258.0 & $1.23\pm0.13$ & $12.4_{-0.4}^{+0.2}$ & $420^{+260}_{-240}$ & $8.64\pm0.04$\\ 
RUBIES-62812  & 673.0 & $1.5\pm0.3$ & $\dagger$$12.46_{-0.21}^{+0.12}$ & $430^{+140}_{-160}$ & $8.88\pm0.07$\\
RUBIES-62012 & 673.1 & $1.2\pm0.3$ & $12.5_{-0.3}^{+0.2}$ & $480^{+250}_{-240}$ & $8.76\pm0.09$\\
\enddata
\tablecomments{Columns: (1) RUBIES ID, (2) AS2UDS ID \citep{2019stach,2020dudz}, (3) ALMA Band 7 flux density reported in AS2UDS, (4) total IR luminosity, (5) star formation rate (SFR) derived from IR luminosity, and (6) dust mass derived from Band 7 flux density, with errors derived from propagating observed flux density error. $\dagger$ IR luminosity for this source is adopted from the AS2UDS catalog \citep{2020dudz}; here, SFR$_{\rm IR}$ is re-derived directly from this measurement. The dearth of photometric detections in the far-infrared ($N=2$) for this source prevents robust fitting using \texttt{MCIRSED}.}
\end{deluxetable}

\subsection{Optical/Near-infrared Continuum Properties}

To characterize the rest-frame OIR component of each galaxy's SED, we model any available HST+JWST photometry simultaneously with the PRISM spectra using the Bayesian SED fitting code, \texttt{Prospector} \citep{2017leja,2017johnson,2021johnson}, following a similar procedure as in \citet{2024degraaffa} and \citet{2024weibelb}. Briefly, we adopt the stellar population synthesis models from the Flexible Stellar Population Synthesis (FSPS) package \citep{2009conroy,2010conroy}, with the MILES spectral library \citep{2006sanchez-blazquez}, MIST isochrones \citep{2016choi,2016dotter} and a Chabrier initial mass function \citep{2003chabrier}. Before fitting, all models are convolved to the PRISM resolution using the line spread function, multiplied by a factor of 1.3 \citep[e.g.][]{2023curtislake, 2024degraaffa}. We flux calibrate the spectrum to the photometry (primarily correcting for slit losses) using the \texttt{Prospector PolySpecFit} procedure and polynomial order $n=2$. For each source, we fix the redshift to the spectroscopic redshift. We assume a non-parametric star formation history (SFH) prior \citep[as described in][]{2019leja}, with a 10-bin model utilizing the \texttt{Prospector} continuity prior. The first 100 Myr of star formation are divided into three bins of 5, 25, and 75\,Myr widths respectively to finely sample the most recent SFH, with the remaining age of the universe filled with seven logarithmically spaced age bins ranging from $\sim$60 Myr to 1 Gyr in length. We adopt a two-parameter \citet{2013kriek} dust law allowing the attenuation around both old ($t > 10$\,Myr) and young ($t < 10$\,Myr) stars to vary over $\tau$ of [0, 2.5] (with the younger stars experiencing both dust screens), and a free dust index $\delta$ to vary over [-1, 0.4] that multiplicatively varies the \cite{2000calzetti} dust law. The stellar metallicity is fit as a free parameter with a logarithmically sampled uniform prior over [0.1\,$Z_\odot$, 1.55\,$Z_\odot$] (the full range of the \citealt{2006sanchez-blazquez} library). We fit with nebular emission tied to the star formation history, with the gas metallicity tied to the stellar metallicity and the nebular ionization parameter $\log (U)$ free in the range [-4,-1]. Both the nebular and continuum intrinsic velocity dispersion are left free in the range [0,1000]\,km\,s$^{-1}$ to account for uncertainty in the PRISM line spread function.

\begin{figure*}
    \centering
    \includegraphics[angle=0,trim=0in 0in 0in 0in, clip, width=1.\textwidth]{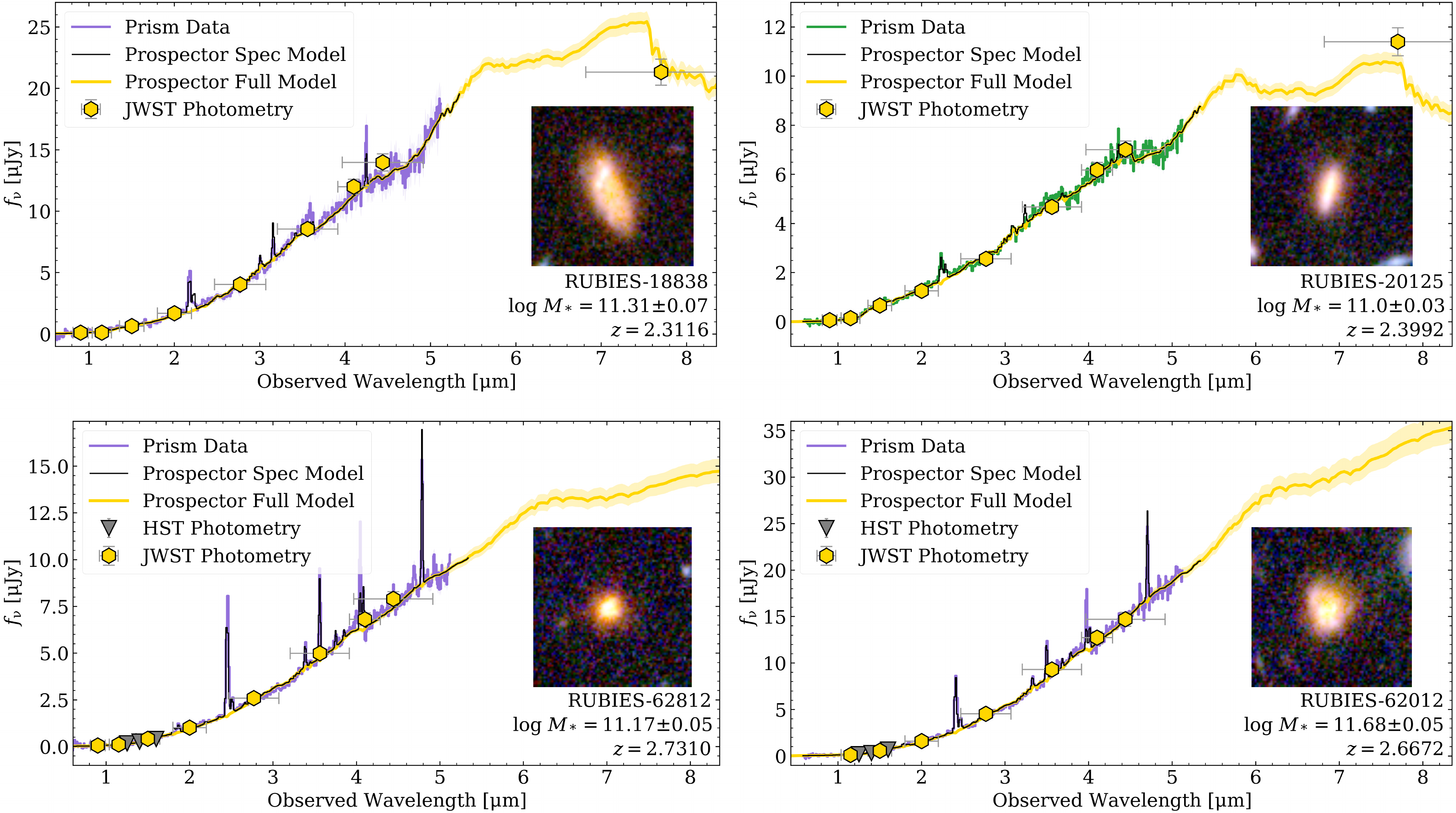}
    \caption{Observed JWST+HST photometry as available, along with median best-fit \texttt{Prospector} model (yellow and black lines). Object names, measured spectroscopic redshifts, and estimated stellar masses from the \texttt{Prospector} fit are listed for each source. The model is fit to the spectra (in purple/green, shown here with slit loss flux calibration applied) and photometry (as gray/yellow points) simultaneously. The JWST/NIRCam RGB cutout derived from F150W/F277W/F444W for each source is shown as an inset ($2''$ wide), with matched scaling across the sample. We present the model SEDs here using a single color rather than adopting the green/purple notation, to emphasize that the multi-wavelength photometry are indistinguishable across the sample.}
    \label{fig:seds}
\end{figure*}

From the median posterior \texttt{Prospector} SEDs, we measure stellar mass, SFR$_{\rm OIR}$, and $A_V$. Note that the $A_V$ derived here is based on SED modeling, and therefore takes into account both the stellar continuum and the nebular emission; in contrast, we measure the recombination lines independently in our spectroscopically-derived $A_V$, as reported in Table \ref{tab:spec}. Similarly, the SFR$_{\rm OIR}$ here is based on rest-frame ultraviolet to near-infrared SED modeling (averaged over the past 10\,Myr, comparable to the timescale of star formation probed by \Ha), informed by both the stellar continuum and emission lines; in Table \ref{tab:spec} we report the \Ha\ and \Pab-based SFRs, which are derived from emission lines independently of the full SED. Results from the \texttt{Prospector} modeling, which includes both the spectroscopic and photometric data, are presented in Figure \ref{fig:seds} and Table \ref{tab:oir}.

\begin{deluxetable}{lccc}
\setlength{\tabcolsep}{0.11in}
\tablewidth{1.\textwidth}
\tabletypesize{\footnotesize}
\tablecaption{OIR-Derived Galaxy Properties\label{tab:oir}} 
\tablehead{
\colhead{Name} & {log$_{10}(M_\star)$} & {SFR$_{\rm OIR}$} & {$A_V$}  \\
	\colhead{}  & \colhead{\Msun}  & \colhead{\Msunyr} & \colhead{ABmag}  \\
	\colhead{(1)}  & \colhead{(2)} & \colhead{(3)} & \colhead{(4)} } 
\startdata
RUBIES-18838 & $11.31^{+0.07}_{-0.08}$ & $110^{+50}_{-20}$ & $3.3^{+0.4}_{-0.3}$ \\ 
RUBIES-20125 & $11.00^{+0.03}_{-0.03}$ & $46^{+10}_{-8}$ & $3.4^{+0.2}_{-0.3}$ \\ 
RUBIES-62812 & $11.68^{+0.05}_{-0.03}$ & $189^{+14}_{-13}$ & $3.08^{+0.07}_{-0.08}$ \\ 
RUBIES-62012 & $11.17^{+0.05}_{-0.05}$ & $380^{+40}_{-30}$ & $3.64^{+0.10}_{-0.10}$ \\ 
\enddata
\tablecomments{Columns: (1) Object ID, with (2) stellar mass ($M_\star$), (3) star formation rate (SFR), and (4) attenuation $A_V$ as measured from the OIR SED, fit to HST+JWST photometry and PRISM spectra simultaneously. All properties and inner 68\% uncertainties are derived from median posterior \texttt{Prospector} SED fits.}
\end{deluxetable}

\subsection{Multi-wavelength Source Characterization}

Each of the four galaxies in our sample share a few common physical properties, consistent with the overall population of DSFGs at cosmic noon \cite[e.g.][]{2015dacunha,2020dudz}. In particular, based on our SED analyses, each galaxy has a large dust reservoir ($\sim10^{8.5}$\,\Msun) and stellar mass ($M_{\star} \geq 10^{11}\,$\Msun) (see Tables \ref{tab:fir} and \ref{tab:oir}). Given their nature as submillimeter-selected, the galaxies are IR luminous with $L_{\rm IR} \geq 10^{12.4}$\,\Lsun, corresponding to a total IR-based SFR of $\geq350$\,\Msunyr. Our independently derived FIR properties are consistent within uncertainties to those reported by AS2UDS \citep{2020dudz}, which are derived using the energy balance procedure \texttt{MAGPHYS} \citep{2008dacunha,2015dacunha,2019battisti} fit to the full pre-JWST ultraviolet-to-radio SEDs. 

We measure attenuation $A_V$ for the sample two ways: based on the Paschen-Balmer decrement, and via \texttt{Prospector} OIR SED fitting, wherein the spectra and photometry are fit simultaneously. While the attenuation measurements using both methods are consistent within uncertainties for each galaxy in our sample, we find the Paschen-Balmer decrement based $A_V$ to be $\sim0.5$\,ABmag higher than the SED based measurement for the two galaxies with relatively fainter emission lines (RUBIES-18838 \& RUBIES-20125). In these cases, the \texttt{Prospector} model has stronger constraints from the stellar continuum rather than the nebular emission, though it takes both into account. This difference is expected and has been observed in previous literature \citep{1988fanelli,1994calzetti,2000calzetti,2011wild,2014price,2014steidel,2015shivaei,2024barrufet}. The relatively higher attenuation toward ionized regions versus stellar continuum can be attributed to the inhomogeneous distribution of gas and dust: the nebular lines may trace a more highly obscured region of the galaxy, as the radiation from young stars is subject to both the diffuse dust seen by all stars and an additional dust component from their birth clouds \citep[for a review see][]{2020salim}.

Below, we detail individual multi-wavelength physical characterization for our sample on a per galaxy basis.

\subsubsection{RUBIES-18838, or AS2UDS.0258.1}

This galaxy was originally discovered as a SCUBA-2 submillimeter source \citep[UDS.0258; ][]{2017geach}, which was resolved into two galaxies upon ALMA follow-up \citep{2019stach}. The ALMA counterpart to RUBIES-18838 is a moderately bright ($S_{870}$$\sim1$\,mJy) submillimeter galaxy \citep[AS2UDS.0258.1;][]{2020dudz}. 

RUBIES-18838 has a spectroscopic redshift of \zs$ = 2.3116$, confirmed by multi-line identification of \Ha+\Nii, \Oiii, \Hei, \Pab, and multiple sulfur lines; the photometric redshift is in good agreement, with \zp$=2.3_{-0.1}^{+0.2}$. Analysis of the PRISM spectrum reveals strong \Ha\ emission (S/N $\sim23$) with EW$_{\rm H\alpha+[NII]}\sim209\,\rm\AA$, suggesting recent or ongoing star formation activity. Based on the Paschen-Balmer decrement, we measure strong dust attenuation in the optical of $A_V = 3.8 \pm 2.1$; the attenuation measured from OIR SED fitting is $\sim0.5$\,mag higher but consistent within uncertainties. We find a dust-corrected SFR of $\sim220$\,\Msunyr\ from the slit-loss corrected \Ha\ and \Pab\ line luminosities, which is generally consistent with the FIR-based and OIR-based SFRs within uncertainties. 

JWST/NIRCam F444W morphological parameters for our sources are derived by \citet{2024gillman}. For this source, they find a Sérsic index of $n=0.53\pm0.01$, fit to the F444W flux. The galaxy is generally disk-like with a high residual flux fraction (RFF) of the Sérsic fit, due to substructures of concentrated emission. It is fairly extended, with $R_{50}^{\rm F444W} = 4.7\pm0.1$\,kpc, and elongated, with an axis ratio of $b/a=0.380\pm0.001$.

The PRISM spectrum demonstrates complex structure, and due to the alignment of the slit with the extent of the galaxy (see Figure \ref{fig:spec}), the integrated 1D spectrum is multi-component, with spatial variation of star-dust geometry and evolved stellar populations detected across the spatial axis of the 2D spectrum. These features altogether suggest that the galaxy is actively starbursting, with some variation in star formation activity, dust content, and stellar population age across the extent of the disk.

\subsubsection{RUBIES-20125, or AS2UDS.0258.0}

RUBIES-20125 corresponds to the same SCUBA-2 submillimeter source as RUBIES-18838 \citep[UDS.0258; ][]{2017geach}, which was resolved into two galaxies upon ALMA follow-up. The ALMA counterpart to RUBIES-20125 is a moderately bright ($S_{870}$$\sim1$\,mJy) submillimeter galaxy \citep[AS2UDS.0258.0;][]{2020dudz}.

RUBIES-20125 has a spectroscopic redshift of \zs$ = 2.3992$, confirmed by multi-line identification of \Ha+\Nii, He\textsc{I}, and \Pab, in good agreement with the photometric redshift \zp$=2.44_{-0.13}^{+0.16}$. Analysis of the PRISM spectrum reveals moderate \Ha\ emission (S/N $\sim12$) with EW$_{\rm H\alpha+[NII]}\sim63\,\rm\AA$, suggesting some recent or ongoing star formation activity. We detect \Hb\ in absorption, further indicating both high dust attenuation and gas depletion. Anchored to the Paschen-Balmer decrement, we find strong dust attenuation with $A_V = 4.0 \pm 2.2$; the OIR SED-based attenuation measurement is $\sim0.5$\,mag higher but consistent within uncertainties.  We measure a dust-corrected SFR of $\sim50$\,\Msunyr from both \Ha\ and \Pab, as well as from OIR SED fitting, which is approximately an order of magnitude lower than the FIR-based SFR (see \textsection4.2 for further discussion). The spectrum is fairly flat in $f_\lambda$, and shows clear evidence of a Balmer break. These features altogether suggest that the galaxy has ongoing but relatively little star formation (compared to the FIR-derived SFR) plus a fairly evolved stellar population, which may suggest that the galaxy is entering a post-starburst phase, defined here by its relatively low ongoing SFR, large Balmer break, and declining SFH.

Taking JWST/NIRCam F444W morphological parameters from \citet{2024gillman}, this source has a Sérsic index of $n=0.90\pm0.01$ (fit to the F444W flux) and has low RFF of the Sérsic fit, indicating a fairly smooth disk-like structure. It has a size of $R_{50}^{\rm F444W} = 2.477\pm0.001$\,kpc, close to the median size \citet{2024gillman} find for their sample of DSFGs. It is elongated, with an axis ratio of $b/a=0.331\pm0.001$.

Based on the source's 1.4\,GHz radio flux density of $50\pm8$\,$\mu$Jy and the total IR luminosity, we calculate the infrared-to-radio luminosity ratio, $q_{\rm IR} = 2.08\pm0.07$. This is lower than the expected value for star-forming galaxies at this redshift, and indicates that the source has radio excess due to some contribution from an AGN \citep[see][]{2017delhaize}. This is supported by the detection of \Feii\ in the spectrum, though we cannot rule out that the emission is induced by shocks. Nonetheless, we caution that the derived OIR-based SFR (both from the emission lines and the full OIR SED) and stellar mass for this source may be overestimated, as the relative AGN contribution to the luminosity of each tracer is not accounted for.

\subsubsection{RUBIES-62812, or AS2UDS.0673.0}

This galaxy was originally discovered as a SCUBA-2 submillimeter source \citep[UDS.0673.0; ][]{2017geach}, and was resolved into two galaxies upon ALMA follow-up. The ALMA counterpart to RUBIES-62812 is a $S_{870}$$\sim1.5$\,mJy submillimeter galaxy \citep[AS2UDS.0673.0;][]{2020dudz}, and is the far-infrared-brightest source in the sample.

RUBIES-62812 has a spectroscopic redshift of \zs$ = 2.7310$, and features many strong lines including \Oiii, multiple sulfur lines, \Ha+\Nii, and multiple Paschen lines. The photometric redshift is somewhat discrepant, with \zp$=2.3_{-0.1}^{+0.4}$. The PRISM spectrum shows strong \Ha\ emission (S/N $\sim42$) with EW$_{\rm H\alpha+[NII]}\sim321\,\rm\AA$, suggesting considerable ongoing star formation activity. The source has strong dust attenuation based on the Paschen-Balmer decrement, with $A_V = 3.1 \pm 2.1$, and is consistent with the OIR SED attenuation measurement within uncertainties. We find a dust-corrected SFR of $\sim320$\,\Msunyr from both \Ha\ and \Pab\, which agrees with the FIR-derived and OIR SED-based SFRs within uncertainties. The spectrum is rising and red in $f_\lambda$. These features altogether suggest that the galaxy is highly star-forming and dusty.

The Sérsic index of this source fit to its F444W flux is $n=0.55\pm0.01$ \citep{2024gillman}. It is very compact with $R_{50}^{\rm F444W} = 1.372\pm0.006$\,kpc, and is generally disk-like with moderate RFF due to some substructures of concentrated emission \citep{2024gillman}. It has a higher axis ratio of $b/a=0.780\pm0.003$.

\subsubsection{RUBIES-62012, or AS2UDS.0673.1}

RUBIES-62012 corresponds to the same SCUBA-2 submillimeter source as RUBIES-62812 \citep[UDS.0673; ][]{2017geach}, and was resolved into two galaxies upon ALMA follow-up. The ALMA counterpart to RUBIES-62012 is a $S_{870}$$\sim1$\,mJy submillimeter galaxy \citep[AS2UDS.0673.1;][]{2020dudz}.

RUBIES-62012 has a spectroscopic redshift of \zs$ = 2.6672$, and has the largest number of detected lines in our sample, featuring emission lines including \Oiii, multiple sulfur lines, \Ha+\Nii, \Hb, \Oii, and multiple Paschen lines; the photometric redshift (\zp$=2.6_{-0.4}^{+0.1}$) is in agreement with the spectroscopic redshift. The PRISM spectrum shows strong \Ha\ emission (S/N $\sim34$) with EW$_{\rm H\alpha+[NII]}\sim204\,\rm\AA$, suggesting significant ongoing star formation activity. While the \Hb\ line is detected at $4\sigma$, we base our measurement of optical dust attenuation on the Paschen-Balmer decrement to be consistent with our measurements for the remainder of our sample, and find a high value of $A_V = 3.6 \pm 2.1$, well matched with the OIR SED-based $A_V$. We find a dust-corrected SFR of $\sim290$\,\Msunyr from both \Ha\ and \Pab, consistent within uncertainty with the FIR- and OIR-based SFRs. The spectrum is rising and red in $f_\lambda$. Similarly to RUBIES-62812, these features altogether suggest that the galaxy is highly star-forming and dusty.

The Sérsic index of this source fit to its F444W flux is $n=1.53\pm0.01$ \citep{2024gillman}. Its size is typical of the \citet{2024gillman} sample, with $R_{50}^{\rm F444W} = 2.115\pm0.004$\,kpc. It has a high RFF, indicating substructures of concentrated emission, and a high axis ratio of $b/a=0.832\pm0.004$.

Based on the source's radio flux density of $48\pm17$\,$\mu$Jy, and the total IR luminosity, we find a low infrared-to-radio luminosity ratio, $q_{\rm IR} = 2.05\pm0.15$, indicating some contribution from an AGN. The \Feii\ emission line in the PRISM spectrum may also be caused by AGN, or shocks. As with RUBIES-20125, the derived SFR and stellar mass for this source may be overestimated, as the relative AGN contribution to the luminosity of each tracer is not accounted for.

\section{Discussion}

\subsection{Steep dust attenuation curve with high $A_V$}

The dust attenuation curve describes the wavelength-dependent efficacy of absorption and scattering from interstellar dust intercepting starlight, that effectively dims shorter-wavelength light more strongly than longer-wavelength light. By applying a dust attenuation curve to an observed galaxy spectrum or SED, the emitted light from the galaxy can be reproduced, in order to accurately determine its dust-corrected physical properties. Therefore, one's choice of dust attenuation curve has a significant impact on key derived properties such as stellar mass and SFR \citep{2013kriek,2015reddy,2015shivaei,2016salim}.

Galaxies spanning a range of redshifts have been shown to exhibit a wide range of attenuation curve shapes and slopes \citep[for a recent review see][]{2020salim}. Commonly adopted curves include the Calzetti curve for local starburst galaxies \citep{2000calzetti}, the Galactic extinction curve \citep{1989cardelli}, and the LMC/SMC extinction curves \citep{2003gordon}. Variation in the steepness of the attenuation curve can arise from changes in the dust grain-size distribution \citep[e.g.][]{2011wuyts}, or from geometrical effects, namely differences in the dust optical depths across the galaxy \citep[e.g.][]{2005inoue,2013chevallard,2024vijayan}. 

Both observational works \citep[e.g.][]{2016salmon,2018salim} and theoretical models \citep[e.g.][]{2013chevallard} have demonstrated the strong correlation between the attenuation curve slope and effective optical opacities, wherein steeper slopes tend to arise from galaxies with a lower dust column density, or in other words, a lower $A_V$ \citep{2020salim}. Indeed, dusty, IR-luminous galaxies like DSFGs with characteristically high $A_V$ are expected to have gray, shallow attenuation curves \citep{2013chevallard,2017leja}. DSFGs are also thought to have complex star-dust geometries; a complex, clumpier star-dust geometry also flattens the attenuation curve, due to the relative ease of short-wavelength photon escape within more complex geometries \citep{2018narayanan}. 

\begin{figure}[h]
    \centering
    \includegraphics[angle=0,trim=0in 0in 0in 0in, clip, width=0.5\textwidth]{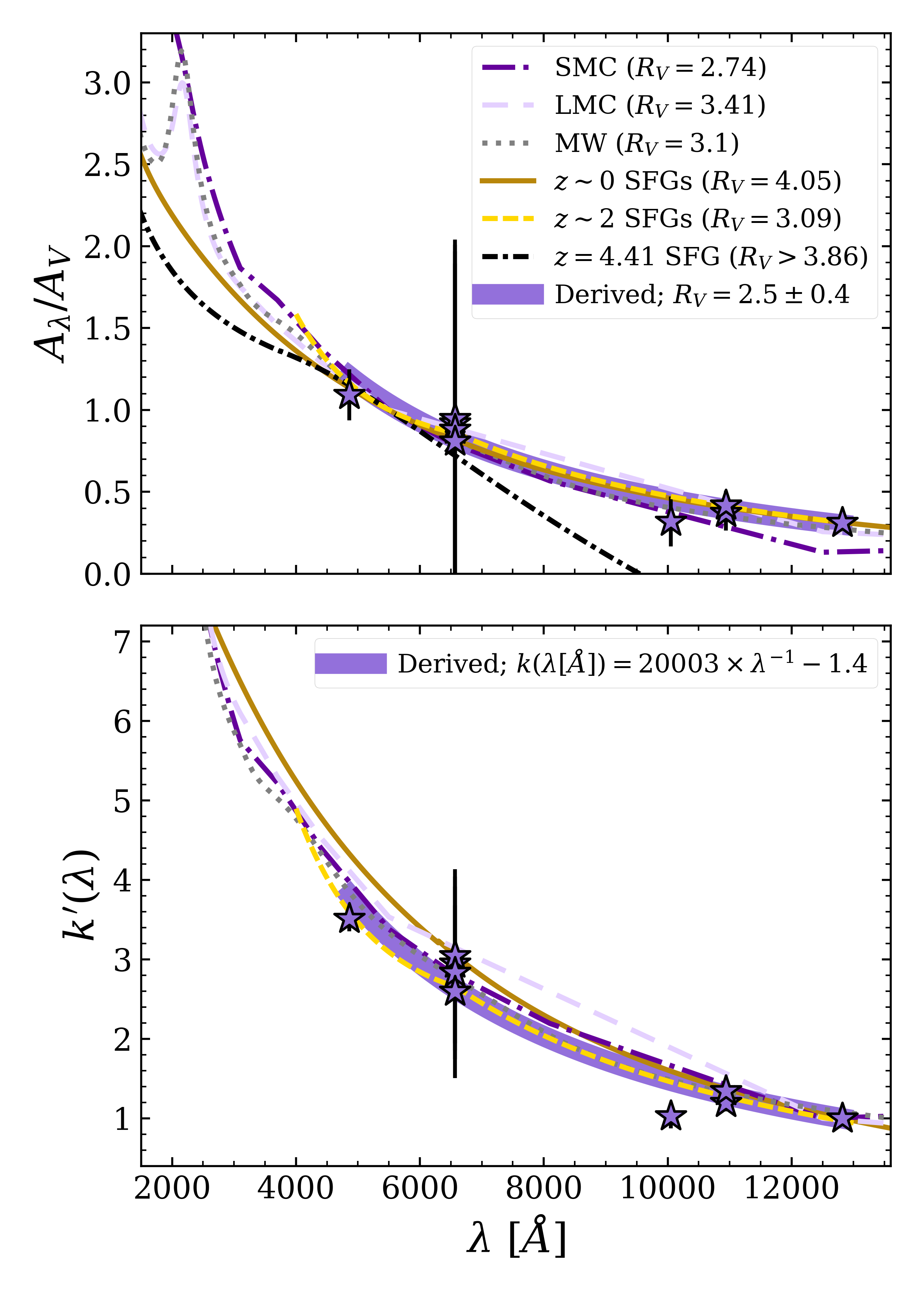}
    \caption{Dust attenuation curves derived for our sample compared to a few literature attenuation/extinction curves. Attenuation curve measurements and associated errors from line ratios measured for all detected Balmer/Paschen emission line pairs are shown as stars, and the best-fit function linear in $1/\lambda$ is shown in thick purple. Overplotted are the extinction curves for the MW \citep[gray;][]{1989cardelli}, SMC \citep[dark purple;][]{2003gordon}, and LMC \citep[lavender;][]{2003gordon}, as well as attenuation curves measured for star-forming galaxies (SFGs) nearby \citep[dark gold;][]{2000calzetti}, at $z\sim 2$ \citep[yellow;][]{2020reddy}, and for a $z=4.41$ galaxy \citep[black;][]{2024sanders}. \textit{Top:} Absolute attenuation curve normalized to $A_V = 1$, for direct comparison of the relative shapes of the attenuation curves. \textit{Bottom:} Total attenuation $k^\prime(\lambda)$, to demonstrate the relative magnitudes of attenuation over the wavelength range. Line ratios used for fitting are all anchored to \Pab, therefore each attenuation curve here is normalized such that $k^\prime(\rm Pa\beta)=1$. Note that we do not show the $z=4.41$ curve here as it is not constrained out to our normalization wavelength. The legend shows the best fit functional form for $k(\lambda)$; this is the normalized functional form such that $k(28000\rm{\AA}) = 0$, which differs in normalization constant only to the curve shown in the figure.}
    \label{fig:alaw}
\end{figure}

As described in \textsection3.2, we derive the normalized attenuation curve $k(\lambda)$ from all Paschen/Balmer series emission lines detected at S/N $>3$ in our sample. Note that this empirical measurement is anchored to attenuated nebular emission, which serves as a proxy for the stellar continuum attenuation; on average, the reddening of nebular lines may be greater than that of the stellar continuum, primarily as a consequence of star-dust geometry towards ionized regions surrounding young stars \citep{2020reddy}. In Figure \ref{fig:alaw}, we present two forms of the attenuation curves. The absolute attenuation (normalized by $A_V$) allows for direct interpretation of the slope of the curve \citep{2018salim}. Here, the different curves vary more in the ultraviolet, wherein the SMC/LMC curves become much steeper than the Calzetti and Milky Way (MW) curves. Note that we present both attenuation (Calzetti and other starburst galaxy derivations from the literature) and extinction (MW, SMC, and LMC) curves, and in this work, we derive an attenuation curve. While extinction is limited to the effects of absorption and scattering, attenuation also incorporates geometric effects of the star-dust geometry within galaxies. In contrast to the absolute attenuation (upper panel of Figure \ref{fig:alaw}), the total attenuation curve $k(\lambda)$ (lower panel) is normalized by the color excess, $E(B-V)$, making interpretation of curve steepness less intuitive but providing a better demonstration of the total attenuation over the wavelength range. Here, we plot $k^\prime(\lambda)$, which is anchored to \Pab, with each attenuation curve normalized such that $k^\prime(\rm Pa\beta)=1$.

We find a Calzetti-like shape of the rest-frame optical/near-infrared attenuation curve, a steeper slope at bluer wavelengths ($\lesssim5000\,\rm{\AA}$) more aligned with the SMC extinction curve, and a very high $A_V$ of $\sim 3-4$\,ABmag for our sample. The dust grain physical properties (i.e., grain size distribution and composition) has a degenerate impact on the shape of the attenuation curve with geometric and optical depth effects. \citet{2020zelko} predict that larger dust grains are correlated with a higher ratio of total to selective optical extinction ($R_V$), which is equivalent to $k_V$. We find a value of $R_V=2.5\pm0.4$, comparable to that of the SMC extinction curve and smaller than that of the Calzetti curve ($R_V=4.05$). The derived $R_V$ is consistent with results from similar observational studies of star-forming galaxies at cosmic noon \citep[e.g.][]{2015reddy,2020reddy}. Notably, \citet{2015reddy} find an $R_V=2.51$ for their sample of 224 star-forming galaxies at $z\sim2$ measured via Balmer lines observed with deep ground-based spectra. This relatively lower $R_V$ compared to the standard starburst model (the Calzetti curve) may indicate differences in the dust grain size distribution, namely smaller dust grains in these higher-redshift environments, or differences in star-dust geometry. The dust grain distribution has been observed to impact the heating and cooling of the ISM \citep{1994bakes,2001weingartner,2020mckinney,2021mckinney}, and recent modeling work has demonstrated large variations in star-formation efficiency when the dust grain size shifts, wherein a smaller dust grain size distribution is correlated with a higher star formation efficiency \citep{2024soliman}. However, as we lack emission lines blueward of \Hb,  our constraints are limited at these wavelengths, and higher order Balmer lines are needed to improve our measurement of the rest-frame optical attenuation curve.

\citet{2024sanders} recently derived the attenuation curve for a star-forming galaxy at $z=4.41$ with deep JWST/NIRSpec observations of 11 unblended recombination lines; this is the highest-redshift attenuation curve measurement to-date. Their derived attenuation curve has a steeper slope at long wavelengths ($\lambda >5000\,\rm\AA$), and a similar slope to Calzetti, SMC, and MW curves in the rest-ultraviolet to optical. Given this difference in shape compared to standard literature curves, \citet{2024sanders} suggest that dust grain properties, star-dust geometry, or both differ from what is observed for local star-forming galaxies. The attenuation curve derived herein suggests a similar conclusion, though importantly our recombination lines are blended, and do not probe the rest-ultraviolet, therefore we lack constraining power to infer the dust properties in greater detail.

Indeed, one important caveat to this finding is that we cannot reliably correct for the contribution of \Nii\ to our measurement of the blended \Ha+\Nii\ emission line flux. This is especially relevant given the AGN signatures present for half of our sample, based primarily on their moderate radio excess. We briefly assess the impact of significant \Nii\ contribution to our findings regarding the dust attenuation properties of our sources. If \Nii\ contributes 50\% of the \Ha+\Nii\ emission line flux, this would increase the value of $A_V$ by about $\sim1.5$\,mag for each source. However, given constraints provided by the other Paschen/Balmer emission lines, the impact on the normalized shape of the dust attenuation curve is negligible. In our derivation, which is also reflected in Figure \ref{fig:alaw}, we include a \Nii\ fraction term of $0.5\pm0.5$, to allow for uncertainty in the contribution from \Nii.

Given the unresolved \Ha+\Nii\ emission in our spectra, the lack of constraints in the rest-frame ultraviolet blueward of \Hb, and the small sample size, this attenuation curve derivation serves only as preliminary efforts using rest-frame optical spectra to constrain the conditions of dust in DSFGs at cosmic noon. Future work to measure dust attenuation curves for larger samples of both DSFGs and broader samples of galaxies at similar redshifts will further our understanding of the nature of dust at an epoch and for galaxy types that until now have been largely inaccessible at this level of detail. Obtaining higher resolution spectra to disentangle the contribution from \Nii, bluer wavelength coverage of higher-order Balmer series lines, and spatially resolved spectroscopic or medium/narrow-band photometric data would each help disentangle the various degeneracies impacting the shape of the attenuation curve. Robust attenuation curve analyses with such data as a function of galaxy properties such as redshift, SFR, stellar mass, and metallicity will enable characterization of star-dust geometry as well as the composition and life cycle of dust in galaxies as they evolve.

\subsection{Spectroscopic confirmation of evolving star formation properties through the DSFG phase}

\begin{figure}[h]
    \centering
    \includegraphics[angle=0,trim=0in 0in 0in 0in, clip, width=0.5\textwidth]{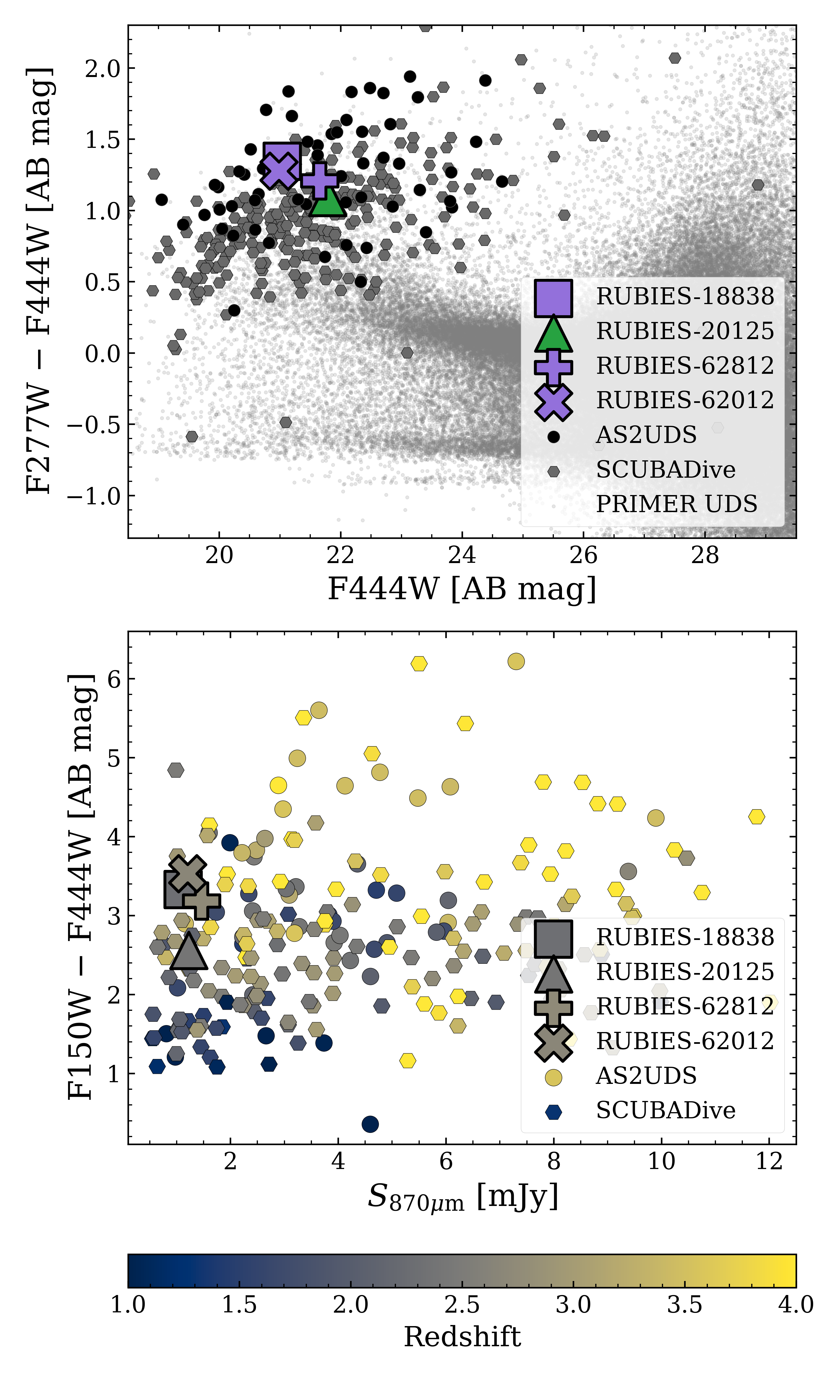}
    \caption{Our DSFGs are bright and red, have typical NIRCam colors and moderately bright submillimeter flux densities, overall consistent with the larger DSFG sample. \textit{Top:} JWST/NIRCam color magnitude diagram comparing the F277W $-$ F444W colors and F444W magnitudes of the DSFGs reported in this work to the broader sample of known DSFGs in AS2UDS \citep[][]{2020dudz,2024gillman} within the PRIMER-UDS footprint (black points), and to the DSFGs in SCUBADive \citep[gray hexagons;][]{2024mckinney}, with the full PRIMER-UDS catalog in the background (small gray points). The three starbursting galaxies in our sample are noted in purple, while the post-starburst galaxy is shown in green. \textit{Bottom:} Color magnitude diagram with JWST/NIRCam F150W $-$ F444W color versus ALMA Band 7 flux density. Symbols are consistent as in the top subplot, but here colors denote the redshift; a spectroscopic redshift for the four RUBIES DSFGs, else the best-available redshifts from AS2UDS and SCUBADive.}
    \label{fig:cmd}
\end{figure}

In the absence of spectroscopic information, all four DSFGs presented in this paper are essentially indistinguishable via their photometry alone, even across multiple energy regimes. Each has been classified in the literature as a typical, starbursting, dusty galaxy at cosmic noon by their ultraviolet-to-radio SEDs \citep[][]{2020dudz}. Even with the addition of deeper, longer wavelength photometry from JWST/NIRCam \citep[e.g.][]{2024gillman}, the galaxies have very similar colors and derived stellar properties. As shown in Figure \ref{fig:cmd}, while the sample is small, the four galaxies sit well within the overall scatter in color-magnitude space occupied by the sample of 68 DSFGs in the AS2UDS sample that lie within the PRIMER-UDS footprint as well as the SCUBADive sample \citep[289 DSFGs with joint ALMA and JWST constraints in the COSMOS field;][]{2024mckinney}, with respect to both JWST/NIRCam colors and ALMA flux densities \cite[][]{2019stach,2020dudz,2024mckinney}.

While the photometry indicate that the four DSFGs in this paper have uniform physical properties, NIRSpec/PRISM spectra reveals distinction between the starburst and post-starburst DSFGs. Three of the DSFGs have many strong emission lines including high \Ha\ EWs, indicating ongoing starburst activity. Their SFRs are on the order of 100s\,\Msunyr\ based on both short (\Ha\ and \Pab, $\sim10$\,Myr) and long (FIR, $\sim100$\,Myr) timescale SFR tracers. The post-starburst galaxy (RUBIES-20125) also has multiple but fewer detected emission lines, with significantly weaker \Ha\ emission. Notably, this source has about an order of magnitude lower SFR derived from the short-timescale indicators, \Ha\ and \Pab, as well as a clear Balmer break. This divergence has been observed for nearby post-starburst galaxies; \citet{2024wild} find that the ratio of FIR to \Ha\ luminosity for post-starbursts can be up to $30\times$ greater than that of star-forming galaxies. 

The combination of FIR photometry with OIR spectroscopy for the post-starburst RUBIES-20125 directly resolves, to zeroth order, its SFH. However, we cannot completely rule out the contribution of dust heating from older stellar populations ($\gtrsim100$\,Myr) to the total FIR luminosity \citep[leading to an overestimation of the FIR-based SFR, e.g.][]{2014utomo,2019leja,2019nersesian}. Still, though we do not input the FIR photometry into the \texttt{Prospector} fitting routine, we do find that the predicted FIR output from the \texttt{Prospector} model for RUBIES-20125 is consistent with the observations. The \texttt{Prospector} model predicts a SFH wherein the galaxy hosted a strong starburst $\sim500\,$Myr ago, followed by some residual and ongoing star formation, constrained by the nebular emission. This interpretation requires that the star formation activity captured by the spectrum (which is only partly aligned with the galaxy center) does not have any strong gradients within the galaxy, which could potentially replicate the spectrum and explain the difference between the FIR photometry and OIR spectrum. For example, the core of the galaxy may be dustier and starbursting (therefore, submillimeter-bright), while the outer regions contain evolved stellar populations, and appear more post-starburst. In this case, we would likely observe a color gradient in the OIR photometry, which is not apparent in the JWST/NIRCam imaging.

Assuming the \texttt{Prospector} solution, the SFH for RUBIES-20125 has potential implications for the evolution of galaxies through the DSFG phase: the source is submillimeter-bright, though the peak of the starburst was estimated to be $\sim500$\,Myr ago, a factor of $2-3\times$ longer than the typical gas depletion timescales for DSFGs at cosmic noon, which are based on their ratios of molecular gas supply to SFR and feedback-induced quenching mechanisms \citep[e.g.][]{swinbank2014,2020dudz,2021sun,2022cooper}. The remaining --- and possibly longer-lasting --- FIR emission may in part be sustained by the $\sim50$\,\Msunyr\ ongoing star formation activity (as predicted by the \Ha\ luminosity), and/or may indicate that the duty cycle of DSFGs is longer than predicted. In other words, these DSFGs may therefore be visible as submillimeter-bright galaxies for at least as long or longer than their starburst phase.

The category of submillimeter-selected galaxies has been shown to be physically heterogeneous, primarily observed through their morphological diversity \citep[e.g.][]{2012hayward,2015smolcic,2020jimenezandrade}. This observation suggests that DSFGs may encompass a heterogeneous population of galaxies hosting starburst activity and dust, rather than representing a single, uniform evolutionary phase in the massive galaxy life cycle. A framework for massive dusty galaxy evolution arises naturally from this case study, as these data suggest that detailed OIR spectroscopy can resolve distinct evolutionary phases for what would be classified photometrically as canonical submillimeter-selected DSFGs, wherein the actively starbursting dusty galaxies with many strong emission lines transition into more evolved, post-starburst galaxies later on. Nonetheless, while evolutionary phases may be inferred from this small sample, drawing broader conclusions is not possible in a statistical sense, and spectroscopic analysis for a much larger sample is required to confirm the diversity of evolutionary phases encompassed by the traditional DSFG category.

\section{Summary}

In this paper we present JWST/NIRSpec PRISM spectra for four typical, ALMA-confirmed DSFGs at cosmic noon that were observed through the RUBIES program. We measure and compare star formation rate indicators across energy regimes and timescales, as well as any other accessible tracers of stellar and dust content for our sample. Our main results are as follows:

\begin{itemize}
    \item Anchored to detections of multiple Balmer and Paschen lines across our sample, we derive an attenuation curve and find its shape to be generally consistent with literature attenuation curves derived for star-forming galaxies locally and at $z\sim2$. The derived curve deviates from the Calzetti (local starburst) curve slightly at bluer wavelengths, where it becomes steeper and more consistent with the SMC extinction curve ($R_V = 2.5\pm0.4$), potentially suggesting a smaller dust grain size distribution for higher-redshift galaxies. Still, higher resolution observations and bluer spectroscopic coverage are needed to break degeneracies with dust composition and star-dust geometry.

    \item Limited to only photometric constraints --- even spanning energy regimes --- the four galaxies in our sample appear to be uniform in their star formation characteristics, and are characterized as canonical DSFGs. However, spectroscopic signatures resolve evolutionary phases of the DSFGs: three of our sources are highly star-forming and are likely experiencing ongoing starbursts, while one source exhibits post-starburst spectroscopic features including a significant Balmer break and a declining SFH. 
\end{itemize}

Through analysis of some of the first high quality rest-frame OIR spectra of $z\sim2.5$ DSFGs, this case study provides detailed spectroscopic evidence that the DSFG category encompasses a heterogeneous sample spanning physical properties and evolutionary stages. Future efforts to spectroscopically characterize star formation activity and dust properties for larger JWST/NIRSpec datasets will allow evolutionary phases of massive galaxies to be traced throughout the Universe's transition from the era of gas accretion to the era of gas depletion, towards and beyond cosmic noon.

\section*{Acknowledgements}

O.R.C, C.M.C., and others at UT Austin acknowledge that they work at an institution that sits on indigenous land. The Tonkawa lived in central Texas, and the Comanche and Apache moved through this area. We pay our respects to all the American Indian and Indigenous Peoples and communities who have been or have become a part of these lands and territories in Texas. We are grateful to be able to live, work, collaborate, and learn on this piece of Turtle Island.

This material is based on work supported by the National Science Foundation Graduate Research Fellowship under grant number DGE 2137420.

C.M.C. thanks the National Science Foundation for support through grants AST-1814034 and AST-2009577 as well as the University of Texas at Austin College of Natural Sciences for support; C.M.C. also acknowledges support from the Research Corporation for Science Advancement from a 2019 Cottrell Scholar Award sponsored by IF/THEN, an initiative of Lyda Hill Philanthropies. 

This work has received funding from the Swiss State Secretariat for Education, Research and Innovation (SERI) under contract number MB22.00072.

Some of the data products presented herein were retrieved from the Dawn JWST Archive (DJA). DJA is an initiative of the Cosmic Dawn Center (DAWN), which is funded by the Danish National Research Foundation under grant DNRF140.

This work is partly based on observations made with the NASA/ESA/CSA JWST. The data were obtained from the Mikulski Archive for Space Telescopes at the Space Telescope Science Institute, which is operated by the Association of Universities for Research in Astronomy, Inc., under NASA contract NAS 5-03127 for JWST. These observations are associated with programs JWST-GO-01837 and JWST-GO-04233.

\software{\texttt{astropy} \citep{2013astropycollaboration,2018astropycollaboration,2022astropycollaboration}, \texttt{Prospector} \citep{2017johnson}, \texttt{EAzY} \citep{2008brammer}, \texttt{grizli} \citep{2023grizli}, \texttt{Jupyter} \citep{2016kluyver}, \texttt{matplotlib} \citep{2023caswell}, \texttt{MCIRSED} \citep{2022drew}, \texttt{numpy} \citep{2020harris}, \texttt{scipy} \citep{2020virtanen}, STScI JWST Calibration Pipeline \citep[\href{jwst-pipeline.readthedocs.io}{jwst-pipeline.readthedocs.io;}][]{2023rigby}}

\facilities{ALMA, HST, JWST}

\bibliographystyle{aasjournal}
\bibliography{sfl_rubies}

\end{document}